\documentclass[aps,prl,twocolumn,nowpacs,superscriptaddress,groupedaddress,amsmath,amssymb]{revtex4}
\usepackage {graphicx,epsfig,graphics,color}

\begin{document}

\title{Direct measurements of the penetration depth in a superconducting film using magnetic force microscopy}

\author{E. Nazaretski}
\affiliation{Los Alamos National Laboratory, Los Alamos, NM
87545}
\affiliation{Brookhaven National Laboratory, Upton, NY 11973}
\author{J. P. Thibodaux}
\affiliation{Department of Physics and Astronomy, Louisiana State
University, Baton Rouge, LA 70803}
\author{I. Vekhter}
\affiliation{Department of Physics and Astronomy, Louisiana State
University, Baton Rouge, LA 70803}
\author{L. Civale}
\affiliation{Los Alamos National Laboratory, Los Alamos, NM 87545}
\author{J. D. Thompson}
\affiliation{Los Alamos National Laboratory, Los Alamos, NM 87545}
\author{R. Movshovich}
\affiliation{Los Alamos National Laboratory, Los Alamos, NM 87545}


\begin{abstract}
We report the local measurements of the magnetic penetration depth
$\lambda$ in a superconducting Nb film using magnetic force
microscopy (MFM). We developed a method for quantitative extraction
of the penetration depth from single-parameter simultaneous fits
to the lateral and height profiles of the MFM signal, and
demonstrate that the obtained value is in excellent agreement with
that obtained from the bulk magnetization measurements.
\end{abstract}

\maketitle

A fundamental property of superconductors is the ability to expel an external magnetic field (Meissner effect), which is screened on the scale of the magnetic penetration depth, $\lambda$. In type II superconductors above a lower critical field ($H_{c1}$), however, it is energetically favorable to allow partial penetration of  the magnetic field in quantized units called vortices. Each vortex carries one flux quantum, $\Phi_0$, and is surrounded by supercurrents, which, in the dilute vortex limit, decay on the same length scale, $\lambda$. The value of $\lambda$ is related to the density of the superconducting electrons, and its quantitative determination is important for understanding of superconducting materials, symmetry of the superconducting state and underlying mechanism of superconductivity.
\cite{Bonalde 2000,
Prozorov 2000, Hardy 1993, Schopohl 1997} Experimental techniques
that measure $\lambda$ in bulk samples  include {microwave
measurements \cite{WHardy:1998,CPBidinosti:2000},} microstrip
resonators \cite{Langley 1991}, two-coil mutual inductances
\cite{Fiory 1988}, muon-spin-rotation \cite{Sonier 1994} and superconducting quantum
interference device (SQUID) magnetometry. \cite{Civale 1987} Recently, scanning probes have been applied to {in-situ}
measurements of $\lambda$. Scanning SQUID \cite{Hicks 2009} and scanning Hall
microscopy \cite{Bending 1997} measure the vortex magnetic field
above the surface of a sample, but the data analysis needs to take into
account many parameters, and multistep lithography is required for
the optimized performance and sensitivity of both probes.
\cite{Koshnick 2008, Chang 1992, Guikema 2008}

Magnetic Force Microscopy (MFM), {a} well established scanned
probe technique, has been used for imaging and manipulation of
individual vortices in thin films and single crystals \cite{Moser
1995, Volodin 1998, Roseman 2002, Auslaender 2009}, but, until
now, MFM has not been implemented for measurements of the absolute values of $\lambda$
due to complexity of acquired images. Roseman et al. \cite{Roseman
2001} estimated the penetration depth from the width of the
lateral constant height scans of the magnetic force across a
vortex, treating the tip as a point object, and found the value of
$\lambda$ several times higher than that obtained by other
methods. In this Letter, we report the measurements of the
{absolute value of $\lambda$} using the MFM technique. We {use the
knowledge of the size and magnetic properties of the probe
tip to develop a model that allows us to fit the acquired MFM
spectra with the penetration depth $\lambda$ as the only fitting
parameter. We compare obtained best fitting values of $\lambda$,
with those measured in a SQUID magnetometer, and demonstrate that
our experimental and modeling approach allows for the extraction of
the penetration depth from the MFM measurements on an individual vortex.

\begin{figure}[h]
\includegraphics [angle=0,width=8.5cm]{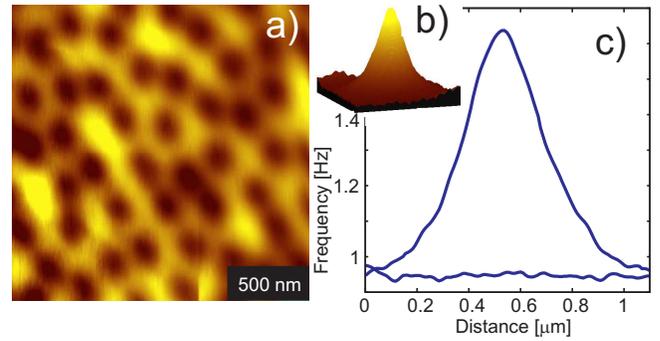}
\caption{(Color online) a) 1.9$\times$1.9 $\mu$m$^2$ scan in a Nb
film at 4.3 K. The field of view contains 53 vortices, and the
vortex density agrees well with the expected value in 30 mT
field. b) image of an individual vortex acquired in the external
field of 1 mT, the color scale corresponds to 0.7 Hz cantilever resonance frequency
change which is a function of the field above the vortex c) cross section of the vortex and the background noise.}
\label{Figure1}
\end{figure}

All measurements described in this Letter were performed in a
home-built low temperature MFM apparatus at T = 4.3 K. Details of
the experimental setup can be found in Ref. \cite{Nazaretski
2009}. We imaged vortices in a 300$\pm$5 nm thick niobium film
fabricated by electron beam deposition. With a tip-sample distance
of $\sim$ 25 $\mu$m, the sample was cooled through T$_c$ (T$_c$ =
8.6 K measured with the SQUID magnetometer) to 4.3 K in the presence
of a magnetic field. The tip was then brought close to
the sample for imaging in the dynamic MFM mode. We used high
resolution SSS-QMFMR cantilever with a resonant frequency of
$f_0\sim 70$ kHz and a spring constant of $c=1.6$ N/m available
from Nanosensors Inc. (www.nanosensors.com). Fig.~\ref{Figure1}
demonstrates imaging of vortices in the Nb film with panel a)
taken in the sample cooled in a field of 30 mT, and panel b)
giving the profile of an individual vortex imaged in the field of
1 mT (yielding large intervortex spacing). It is important to
mention that standard MFM cantilevers have difficulties resolving
individual vortices in fields higher than 10 mT
 \cite{Volodin 1998} and an ultra-sharp double-pyramid design of MFM tips has been used in our probes.
 Panel c) shows the cross section of the vortex, and is a typical curve used for
fitting below. We also recorded MFM spectra as a function of the probe-sample separation.

The measured frequency shift,
\begin{equation}
\Delta f = \frac{f_0}{2c} \frac{\partial F_z}{\partial z}\,,
\end{equation}
depends on the force on the cantilever due to the spatially
varying magnetic field of the vortex. At each point with a local
magnetic moment $\mathbf{m}(\mathbf{r})$, the force
$\mathbf{F}(\mathbf r)=\mathbf{\nabla}[\mathbf{m(\mathbf{
r})}\cdot\mathbf{B(r)}]$, so that the net force on the tip is
\begin{equation}
  \mathbf{F} =\int_{tip}\mathbf{\nabla}[\mathbf{m(
r)}\cdot\mathbf{B(r)}] d\mathbf{r}\,.
\end{equation}
Previous work showed that, if the tip is treated as a point
object, the resulting value of the penetration depth is nearly two
times larger than expected \cite{Roseman 2001}. Consequently, we
parameterize the tip as two cones as shown in the inset of
Fig.~\ref{f:modelresult} d), based on scanning electron microscope
images of a tip. The tip is covered with a magnetic Co-Cr film of
thickness {12 nm}, and the film is saturated in an external field
prior to measurements to the saturation magnetization of
{$M_{sat}\approx 0.8$ T}~\cite{Takahashi 1991}, which gives the
magnetic moment per unit volume of {$6.4\times 10^5$ A/m}.

\begin{figure}
\includegraphics{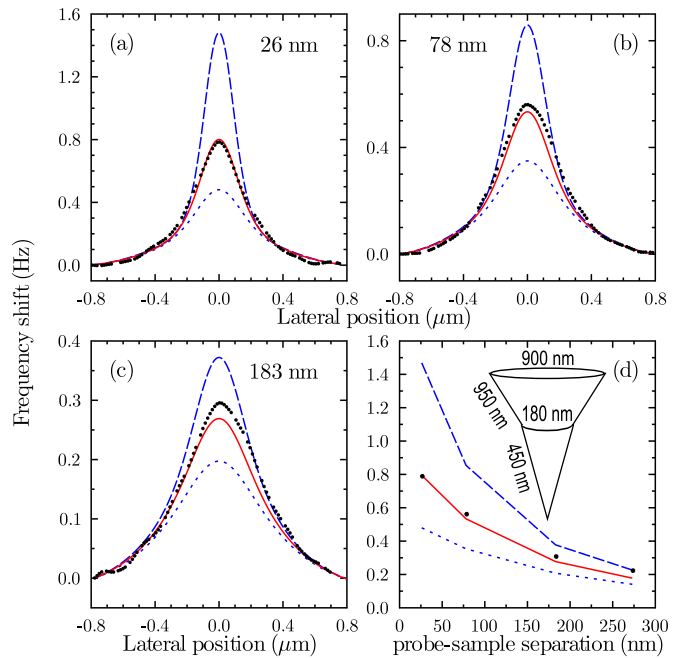}
\caption{\label{f:modelresult}(Color online) Comparison of the model
to the experimental data (dots) for lateral scans, a)-c), and the
probe-sample separation (vertical) scans, d). $\lambda=60,109,160$ nm correspond
to dotted, solid, and the dashed lines respectively. The
probe-sample distance is indicated in each panel.}
\end{figure}

Since the thickness of the superconducting sample is several times
the penetration depth (the assumption confirmed by the value of
$\lambda$ we find below), the field at distance $z$ above the film
surface is very close to that of the vortex outside of a bulk
superconductor, and is very well approximated by the field of a
magnetic monopole of magnitude $2\Phi_0$, where $\Phi_0$ is located at distance $d$ below the
surface~\cite{Carneiro 2000}. The far, $z\gg \lambda$ (near,
$z\leq\lambda$) field is best described by $d=\lambda$
($d=1.27\lambda$) \cite{Carneiro 2000}, and we use the
interpolation formula $d=\lambda(1+0.27/(1+z^2/\lambda^2))$ in our
fitting procedure. Importantly, with this
interpolation, the penetration depth is the only fitting
parameter.

The fits to the MFM scans are shown in Fig.~\ref{f:modelresult},
and demonstrate that $\lambda_f=109$ nm gives a high quality
agreement with the experimental results.  To obtain this value we used the
Levenberg-Marquardt non-linear least squares fitting
algorithm~\cite{NR}, where the variance of the experimental data
relative to the best fit is $\sigma_e^2 = N^{-1} \sum_i \left( f_i
- f_m(x_i,\lambda_f) \right)^2$, where $f_m$ is the model
prediction for a given data point ($x_i$) and $f_i$ is the
experimental data point. Fig.~\ref{f:chisq} shows the measure of
the fit quality, $\chi^2 = ({N\sigma_e^2})^{-1}\sum_i \left( f_i -
f_m(x_i,\lambda_f) \right)^2$, so that $\chi^2 \approx 1$ means a
good fit. The deeper minimum in Fig.~\ref{f:chisq} b) indicates
that the vertical scan is more restrictive for our fitting
procedure. Consequently, the simultaneous fit to both scans yields
$\lambda_f=109\pm 11$ nm, with the variance $\sigma_\lambda^2 =
\sigma_e^2
  \sum_i \left( \frac{\partial f_m(x_i,\lambda)}{\partial \lambda}
  \right)^{-2}$.
This value exceeds the penetration depth for pure Nb,
$\lambda_0=39$ nm, likely due to the presence of disorder, see below.
\begin{figure}
\includegraphics{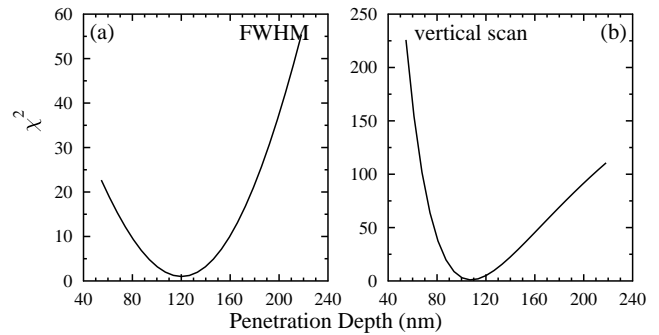}
\caption{\label{f:chisq} Fit quality $\chi^2$ for the full width
half maximum (FWHM) of the a) lateral scan and for the b) vertical scan for 26, 78, 183, and 273 nm probe-sample separations.}
\end{figure}

We used a SQUID magnetometer to verify our result for $\lambda$  by dc magnetization. Measurements were performed on a 5x3 mm$^2$ Nb film with the field applied normal to the surface. A typical magnetization hysteresis loop M(H), obtained at T = 4.5 K after cooling the sample from above $T_c$ in zero field, is shown in Fig.~\ref{f:squid}. $H_{c1}$ cannot be used for a reliable determination of $\lambda$  since demagnetization effects in this configuration are large and the vortex penetration field is strongly increased by pinning. In contrast, the measurement determines the upper critical field, $H_{c2}=\Phi_0/2\pi\xi^2(T)$ (marked by the vertical arrow in Fig.~\ref{f:squid}), where $\xi$  is the superconducting coherence length. We fit $H_{c2}$(T) with a straight line with $T_c =
8.6$ K and the slope $dH_{c2}/dT = -3200$ Oe/K. Using the dirty limit results $\xi(T)=0.855\sqrt{\xi_0 l/(1-T/T_c)}$, and  $\lambda(T)$=$\lambda_0(T)$ $\sqrt{(\xi_0 / 1.33 l)}$ \cite{Tinkham} with the clean Nb values  $\xi_0=38$ nm and  $\lambda_0$ = 39 nm, we obtain the electronic mean free path $\it l $$\approx$ 4.2 nm and $\lambda$(0)$\approx$ 102 nm (thus $\lambda$(4.3K) $\approx$ 105 nm), in excellent agreement with the MFM fits above.


\begin{figure}[h]
\includegraphics[width=0.8\columnwidth]{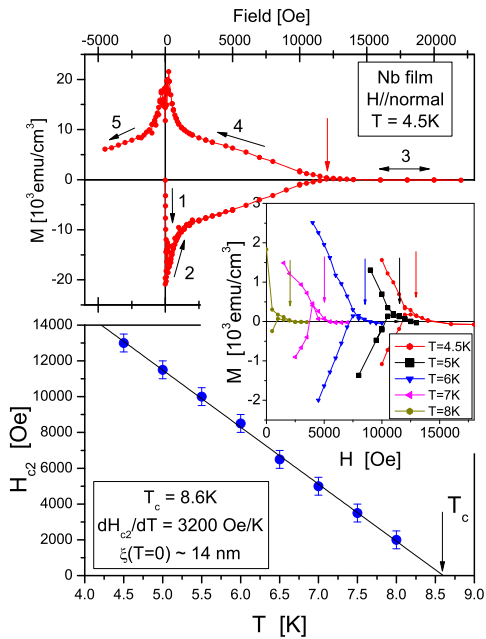}
\caption{\label{f:squid} (Color online) Measurements of the dc
magnetization in the Nb film. Top panel: Hysteresis loop. Segment
1 is the Meissner response, segments 2 and 4 are in the
vortex state, with hysteresis due to pinning. The vortex signal
disappears along segment 3, above the upper critical field
$H_{c2}$. Middle panel: expanded high field region used for
determination of $H_{c2}$ at different temperatures. Bottom panel:
temperature dependence of $H_{c2}(T)$ 
}
\end{figure}

To conclude, we have developed a method for reliable extraction of the
numerical values of $\lambda$ from
MFM images of vortices in superconducting films. We
demonstrated the viability of the method by studying the Nb film.
We expect that this approach will be used for ${\it local}$ determination of the
penetration depth in a variety of novel superconductors and opens
new avenues for the application of MFM in {\em quantitative}
studies of materials.

We acknowledge technical assistance of J. K. Baldwin with
fabrication of Nb films. This work was supported by the US DOE at
Los Alamos and via Grant No. DE-FG02-08ER46492 (I. V. and J. P.
T.),  by the Louisiana Board of Regents (J. P. T.), and was
performed, in part, at the Center for Integrated Nanotechnologies
at Los Alamos and Sandia National Laboratories.

\end{document}